# OmicsQ: A User-Friendly Platform for Interactive Quantitative Omics Data Analysis

Xuan-Tung Trinh[1,a], André Abrantes da Costa[1,2,a], David Bouyssié[3,4], Adelina Rogowska-Wrzesinska[1], Veit Schwämmle[1]

[1] Department of Biochemistry and Molecular Biology, University of Southern Denmark, Campusvej 55, 5230 Odense, Denmark
[2] BioISI – Instituto de Biosistemas e Ciências integrativas, Faculdade de Ciências da Universidade de Lisboa, 1749-016, Lisbon, Portugal
[3] Infrastructure Nationale de Protéomique, ProFI, UAR 2048, Toulouse, France
[4] Institut de Pharmacologie et de Biologie Structurale (IPBS), CNRS, Université de Toulouse (UT), Toulouse, 31077, France
[a] Shared first author

**Abstract**

**Motivation:** High-throughput omics technologies generate complex datasets with thousands of features that are quantified across multiple experimental conditions, but often suffer from incomplete measurements, missing values and individually fluctuating variances. This requires sophisticated analytical methods for accurate, deep and insightful biological interpretations, capable of dealing with a large variety of data properties and different amounts of completeness. Software to handle such data complexity is rare and mostly relies on programming-based environments, limiting accessibility for researchers without computational expertise.

**Results:** We present OmicsQ, an interactive, web-based platform designed to streamline quantitative omics data analysis. OmicsQ integrates established statistical processing tools with an intuitive, browser-based visualization interface. It provides robust batch correction, automated experimental design annotation, and missing-data handling without imputation, which ensures data integrity and avoids artifacts from a priori assumptions. OmicsQ seamlessly interacts with external applications for statistical testing, clustering, analysis of protein complex behavior, and pathway enrichment, offering a comprehensive and flexible workflow from data import to biological interpretation that is broadly applicable tov data from different domains.

**Availability and Implementation:** OmicsQ is implemented in R and R Shiny and is available at https://computproteomics.bmb.sdu.dk/app_direct/OmicsQ. Source code and installation instructions can be found at https://github.com/computproteomics/OmicsQ

**Contact:** veits@bmb.sdu.dk

**Supplementary Information:** Additional documentation and a tutorial are available at Supplementary Material.

# Introduction

The rapid advancements in high-throughput omics technologies have generated an unprecedented volume of quantitative data, necessitating robust and scalable computational solutions for meaningful interpretation [1]. However, challenges such as missing values, batch effects, and proper data normalization complicate downstream analyses and can introduce biases in the biological interpretation if not handled correctly [2].

Missing values are a common issue in omics datasets and can significantly impact statistical inference and biological interpretation. Naive imputation methods may introduce artificial structure into the data and can underestimate variability, potentially inflating false discovery rates and leading to erroneous conclusions [2,3]. Preferably, an optimal analysis avoids imputation altogether, opting instead for statistical methods that operate directly on incomplete data, preserving its intrinsic variability. User-friendly tools for an overall imputation-free analysis are so far missing.

Batch effects, arising from systematic differences between measurements processed in separate experimental batches, can obscure true biological signals if not properly accounted for [4]. While correction methods such as ComBat [5] and the batch correction in limma [5,6] are widely used, care must be taken to avoid overcorrection, especially when batch factors are confounded with biological variables. Although these methods were originally designed for data from microarray experiments, they are generally applicable to quantitative data with batch effects.

Normalization in omics studies is the process of eliminating biases introduced by non-biological, systematic factors - such as technical or experimental variations (e.g., inconsistencies in sample preparation and data acquisition), and adjustment of the normalization method to the biological context (e.g. minimal changes after slight perturbations or vast changes when comparing different tissues). Selecting an appropriate normalization method - whether it is global scaling, LOWESS regression, or ANOVA-based models- requires a careful balance: it should effectively capture unwanted variation without overfitting the data [2].

Given the myriad of tools for quantitative analysis, and particularly their availability as libraries of scripting languages like R and Python, end users without programming skills often rely on simplified solutions that do not implement recent and more powerful methods for data treatment, presenting significant accessibility barriers for experimental researchers, and providing an overall platform to feed their data files with. Moreover, instead of relying on a one-workflow-fits-all solution, end users often prefer to interact with the data to test different hypotheses and/or to confirm *a priori* knowledge about the given biological system. Similar approaches such as integrative omics analysis [7], and web-based solutions providing overall quantitative analysis exist [8], but we believe that they lack modularity and the ability to extensively interrogate the data using modern statistical approaches including variance-sensitive clustering and analysis on protein complex level.

OmicsQ is designed to address these challenges by offering an intuitive, interactive, browser-based platform that streamlines key data-processing steps. By integrating automated experimental design annotation, direct handling of missing values, and interactive quality control visualization, OmicsQ empowers researchers to focus on biological interpretation rather than overcoming computational hurdles. Unlike traditional approaches that rely on imputation to handle missing data, OmicsQ prioritizes statistical techniques that maintain data integrity, reducing the risk of inflated false discovery rates [2]. Furthermore, OmicsQ seamlessly integrates with external applications for statistical testing, clustering, and protein complex analysis, enabling a comprehensive workflow from data import to functional insights. By bridging the gap between computational complexity and usability, OmicsQ provides an accessible and reliable solution for quantitative omics data analysis.

## Methods and Features

OmicsQ was implemented as an interactive, user-friendly interface built with the R Shiny framework, featuring real-time feedback through data summaries, visual projection to principal components and correlation plots. This allows users to directly observe the impact of data-processing choices, and thus should enable informed decisions before continuing with in-depth analysis such as statistical testing and variance-sensitive clustering.

By using an interactive interface with automatic data property detection and extensive help features, OmicsQ aims to provide clear guidance through critical analysis steps, beginning with the file import. Upon uploading Excel or CSV/TSV tables containing features that are quantified across different samples, such as for experimental conditions and replicates, OmicsQ automatically detects table file formats, and permits manual adjustments when necessary. Users then assign specific columns of the uploaded data matrix as identifiers or quantitative measurements for downstream analysis.

Annotation of experimental sample groups and replicates can be a cumbersome task when dealing with many samples. OmicsQ streamlines annotation through semi-automatic comparison of sample names using multiple string-distance metrics (e.g., Levenshtein, Jaro-Winkler). This automatically creates separate groups of samples with similar names from the file header. Users can interactively adjust threshold settings to optimize grouping accuracy rapidly, significantly simplifying experimental design annotation. Samples with consistent naming are easily recognized and grouped with minimal manual effort, reducing the need for time-consuming and error-prone manual annotations across tens or even hundreds of samples.

Advanced pre-processing options include direct handling of missing values without requiring imputation by adding potentially incorrect values to the data set. Batch effects can be corrected using two very common methods from ComBat and limma R packages, and visually evaluated using an integrated principal component analysis (PCA) plot and a correlation heatmap that can be downloaded as pdf files. Furthermore, there are different options for filtering, summarization to main features and normalization. OmicsQ provides a summary of the data set and its main features such as balancing, potential batch effects and dynamic range.

A significant feature of OmicsQ is its integration with specialized statistical analysis applications. These tools are specifically tailored to address the challenges posed by high variability and limited statistical power in omics data, being able to run without requiring data completeness or relying on imputation methods based on stringent assumptions.

The following methods and their Shiny applications have been integrated with OmicsQ:

*PolySTest [9]* is a statistical testing tool tailored for multi-group comparisons in omics data with missing values. It includes Miss Test, which leverages the pattern of missingness itself to detect differentially regulated features without requiring imputation. The integration with OmicsQ enables users to perform statistically rigorous hypothesis testing across different experimental designs while maintaining data fidelity. PolySTest has been updated and is now also available as R package via Bioconductor.

*VSClust [10]* is a variance-sensitive clustering algorithm developed to identify co-regulated molecular features in noisy omics datasets. The main idea of such a procedure is to identify groups of e.g. proteins or genes with similar quantitative behavior between the different experimental conditions, often providing in-depth views into the involved biological pathways and other processes. Its newest versions accommodate missing values by calculating feature distances also for incomplete data, taking advantage that distance calculations are between features and cluster centers, the latter always being complete. Most importantly, VSClust integrates variance estimation into the clustering procedure, improving sensitivity to subtle expression trends and associated biological pathways. Hence, when used via OmicsQ, users can directly explore co-expression patterns and abundance profiles across multiple experimental conditions.

*ComplexBrowser [11]* is an interactive tool for quantifying protein complexes based on subunit co-abundance across samples. It enables users to evaluate complex-level behavior such as stoichiometry changes or coordinated regulation. Through its connection with OmicsQ, users can directly assess known protein complexes following pre-processing and statistical analysis. This feature is restricted to data sets with gene or protein identifiers. OmicsQ automatically converts gene-based IDs like Ensembl IDs and gene names using the UniProt API [12], thus allowing the application of protein complex analysis more generally to e.g. data from RNAseq experiments.

*CoExpresso [13]* can be called from ComplexBrowser to provide a comprehensive view of protein co-expression relationships across human tissues and cell types, based on the ProteomicsDB [14] dataset comprising more than 200 human cell lines or across the thousands of cancer samples from Proteomics Data Commons (PDC) [15]. Submitted protein lists from ComplexBrowser are evaluated whether observed co-regulation patterns are supported by independent biological evidence, supporting hypothesis generation and validation of potential complex formations. The software has been recently updated to include a PDC background with more than 3000 samples and a more recent version of ProteomicsDB, thus considerably increasing statistical power and coverage.

To facilitate seamless interaction among these tools, OmicsQ employs JavaScript messaging to these applications. The default settings call the publicly available instances on [https://computproteomics.bmb.sdu.dk](https://computproteomics.bmb.sdu.dk). Alternatively, one can direct the in-depth analysis to local installations. Docker-based containerization of all five Shiny Apps ensures straightforward local deployment.

With the integration of above applications, users can seamlessly transfer pre-processed data directly to PolySTest for statistical testing, VSClust for variance-based clustering, and ComplexBrowser for protein-complex analysis. When dealing with human data, the general behavior of protein complexes can further be investigated in CoExpresso. Results from statistical testing and clustering can be retrieved and integrated back into OmicsQ, consolidating analyses within a unified environment. Furthermore, OmicsQ furthermore uses the StringDB API [16] to run functional enrichment and visualize the respective protein-protein interaction networks.

With its interactivity and modularity, OmicsQ supports flexible, interactive and user-friendly quantitative analysis applicable to diverse quantitative omics studies, such as proteomics, transcriptomics, metabolomics, and associated data types. Moreover, the software features a tutorial including a representative workflow exemplifying the entire analysis.

## Discussion & Conclusion

OmicsQ offers an interactive and accessible platform that can be operated entirely in a web browser. This lowers the barrier for experimental researchers, clinicians, and students who may lack coding experience but still need to perform high-quality, quantitative omics analyses with confidence. For instance, a biologist analyzing proteomics data from a small cohort of patient samples can quickly upload their dataset, correct batch effects, inspect data variability, and proceed to pathway enrichment—without needing to write a single line of code. Unlike many other tools, OmicsQ does not rely on imputation, and thus preserves data integrity for more reliable downstream analyses, as it does not add potentially incorrect information and lead to misleading results from a significance analysis.

The platform is designed to promote exploratory data analysis, enabling users to iteratively test alternative hypotheses and make informed decisions throughout their workflow. The integration with multiple external applications allows for flexible extension into statistical testing, clustering, and protein complex analysis.

A key strength of OmicsQ is its capacity to maintain a balance between automation and interactivity. Users can inspect each transformation step, review visual summaries (e.g., PCA, heatmaps), and trace parameter choices for enhanced reproducibility. The summary of processing steps ensures that users are aware of every operation performed during the analysis.

Despite its broad applicability, OmicsQ has some limitations. Scalability remains a challenge when dealing with very large datasets, both in terms of memory usage and response time of the interface. The current implementation is optimized for medium-sized datasets (tens to hundreds of samples), typical of many proteomics and transcriptomics studies. Work is ongoing to introduce more efficient data handling and scalable processing beyond simple multi-threading to improve performance.

In summary, OmicsQ provides a robust, user-friendly environment for analyzing quantitative omics data. It empowers researchers to carry out essential preprocessing, hypothesis testing, and functional interpretation tasks without requiring programming skills. By bridging usability with analytical rigor, OmicsQ contributes to more reproducible and insightful omics research workflows.

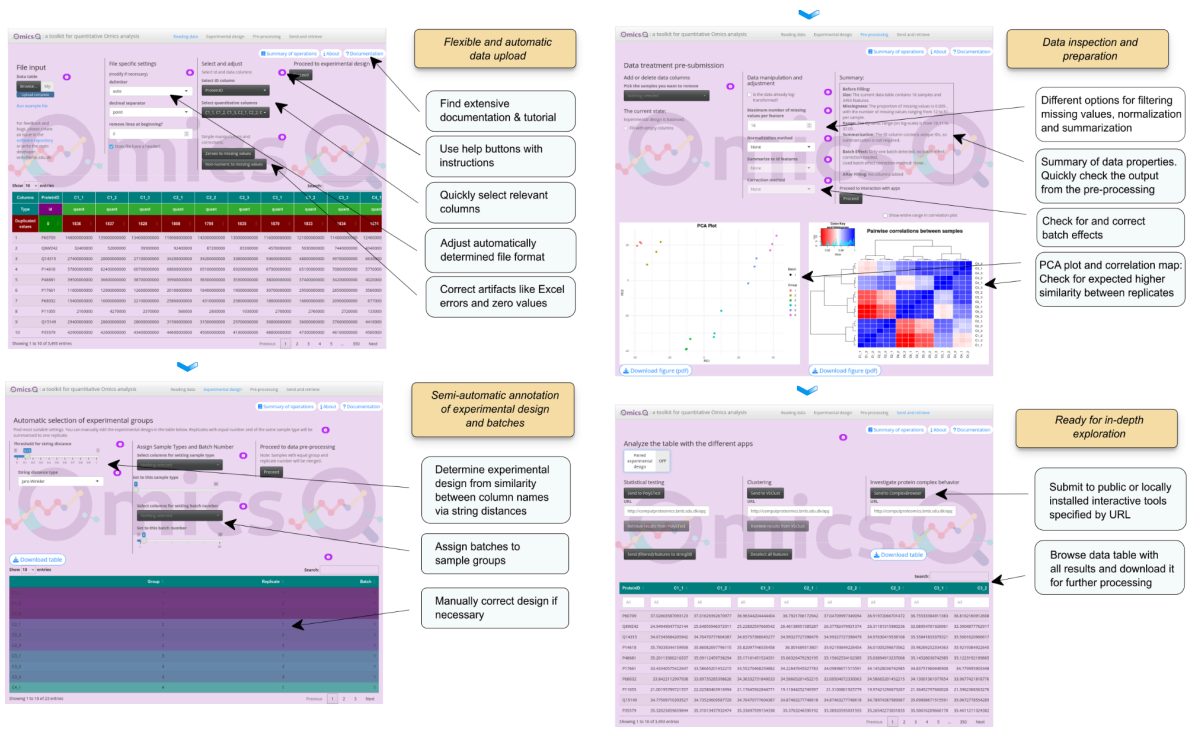

Figure 1: *Overview of OmicsQ features.*
Diagram illustrating the key steps in the OmicsQ workflow: data upload, experimental design annotation, pre-processing (including batch correction and missing value handling) and quality control, and interaction with PolySTest, VSClust and ComplexBrowser.

# Acknowledgements

The authors wish to acknowledge the contributions of the Protein Research Group, Department of Biochemistry and Molecular Biology, University of Southern Denmark, for testing the different versions of OmicsQ during its development phase. AAC was supported by a Fundação para a Ciência e Tecnologia (FCT) PhD fellowship (UI/BD/153051/2022) and by the BioISI Research Unit through FCT grants UIDP/04046/2020 and UIDB/04046/2020.